\documentclass[journal]{IEEEtran}
\IEEEoverridecommandlockouts
\ifCLASSINFOpdf

\else

\fi
\usepackage{amsmath}
\usepackage{graphicx}
\usepackage{subfig}
\usepackage{diagbox}
\usepackage{caption}
\usepackage{epstopdf}
\usepackage{cite}
\usepackage{amsfonts}
\usepackage{amssymb}
\usepackage{algorithmic,algorithm}
\begin{document}	
	\title{Hybrid {C}odeword {P}osition {I}ndex {M}odulation for {S}parse {C}ode {M}ultiple {A}ccess {S}ystem\\
	}
	
	\author{ Ke Lai, Jing Lei, Lei Wen,  Gaojie Chen \IEEEmembership{Senior Member,~IEEE}, Pei Xiao \IEEEmembership{Senior Member,~IEEE} \\ and Amine Maaref \IEEEmembership{Senior Member,~IEEE}
		
	}
	
	\maketitle
	
	\begin{abstract}
		In this paper, a novel variation of codeword position index based sparse code multiple access (CPI-SCMA) system, which is termed as hybrid codeword position index modulated sparse code multiple access (HCPI-SCMA), is proposed to further improve the transmission efficiency (TE). In this scheme, unlike the conventional CPI-SCMA that uses only one kind of bits-to-indices (BTI) mapper, the codeword positions which are padded with zeros in CPI-SCMA are also utilized to transmit additional information. Since multiple index selectors are used in a HCPI-SCMA codeword, the original message passing algorithm (MPA) no longer works in HCPI-SCMA; hence, a modified MPA is proposed to detect the received signals. It is shown in the simulations and analysis that the proposed scheme can achieve both higher TE and better error rate performance in the region of high signal-to-noise ratio (SNR) compare to the conventional SCMA (C-SCMA). Moreover, compared with CPI-SCMA, HCPI-SCMA can achieve higher TE with approximately the same error rate performance compared to CPI-SCMA at high SNRs. 
	\end{abstract}
	
	
\section{Introduction}
In recent years, with the commercialization of the 4th generation (4G) mobile network, researchers started to pay attention to new air-interface technologies which is more efficient and reliable to meet the demand of the 5th generation (5G) mobile network. Non-orthogonal multiple access (NOMA) is proposed as a supplement of the traditional orthogonal multiple access \cite{Andrews2014What}. As a code domain NOMA, SCMA \cite{Hosein2013SCMA} is considered to be a promising 5G candidate under heavily loaded scenarios, such as internet of things (IoT) in massive machine type of communication (mMTC), due to its  ability to support massive users. 

Inspired by the idea of index selection strategy in multi-input multi-output (MIMO) and orthogonal frequency division multiplexing (OFDM) systems, which are termed as spatial modulation (SM) \cite{Mesleh2008SM,Yang2014Design} and OFDM with index modulation (OFDM-IM) \cite{Ba2013Orthogonal,Dang2018adaptive}, respectively. The application of index selection principle to NOMA, especially SCMA has been proposed, which is called CPI-SCMA\cite{LaiCPI}. It should be noted that CPI-SCMA is different from those schemes that directly combine NOMA with SM\cite{Liu2017Spatial,Zhu2017NOMA}. In CPI-SCMA, the information bits of each user not only map to the $M$ point SCMA codebook $\mathcal{S}$, but also map to a BTI mapper which selects the activated positions in a CPI-SCMA codeword. Therefore, the indices of codeword positions in CPI-SCMA carry additional information. Therefore, it is obviously that CPI-SCMA is able to convey more information compared to the conventional SCMA (C-SCMA) with the same number of physical resources, i.e., the TE of CPI-SCMA can be higher than C-SCMA if the parameters are properly selected.

To further improve the TE of CPI-SCMA and the designing flexibility of CPI-SCMA system, in this paper, HCPI-SCMA that fully utilizes the inactivated codeword positions is proposed. In this scheme, a codeword is constructed by $R$ different C-SCMA codebooks and BTI mappers. Based on the original CPI-SCMA, HCPI-SCMA adopt various codebooks to implement CPI-SCMA on the remaining inactivated codeword positions, which indicates that the inactive positions in CPI-SCMA are occupied by SCMA codewords via hybrid index modulation. As different codebooks and BTI mappers are used, the original MPA and message passing aided detection (MPAD) algorithm in CPI-SCMA become inapplicable to the detection of HCPI-SCMA. To tackle this problem, a novel detection algorithm with the aid of merging codebook based MPA (MC-MPA) is introduced. After processing MC-MPA, extra operations to detect the inactive and active indices are applied to the obtained messages. As such, both the indices of activated positions and their corresponding data in HCPI-SCMA can be detected simultaneously. However, it should be noted that the proposed detection algorithm is not yet optimized to achieve maximum likelihood detection performance.

The remainder of this paper is organized as follows. In Section II, the system model considered in this
paper is described. The proposed novel detection algorithms for HCPI-SCMA is discussed in Section III.  The analytical and simulation results are provided in Sections V. Finally, conclusions are drawn in Section VI.

\section{Hybrid codeword position index modulation for SCMA}

\subsection{System model}

We consider the uplink transmissions where $J$ single-antenna users transmit to the same base station (BS) with $K$ allocated resources. In CPI-SCMA, the transmitted $m$ bits for each data block are divided into two parts, the first $m_{1,cpi}$ bits are used to select the active positions from the predefined BTI mapper, and thus
$m_{1,cpi} = \lfloor \log_2(C(n,t)) \rfloor$,
where $C(n,t)$ represents the binomial coefficient and $\lfloor \cdot \rfloor$ denotes the floor function,
and the rest $m_{2,cpi} = t \cdot \log_2C$ bits map to $t$ SCMA codewords, where $t$ is the number of active positions from $n$ available positions. 

Taking the CPI-SCMA system with $n = 4, t = 2$ as an example, only two chosen positions are used to convey SCMA encoded codewords along with the information that the corresponding indices can carry. However, there are still two unoccupied positions that do not transmit data, which is a waste of physical resources. Therefore, utilization of these positions in CPI-SCMA should be taken into consideration. In this paper, in order to further improve the TE, the idea of employing multi-mode codeword position index modulation is introduced in original CPI-SCMA. This scheme is able to improve the TE and sufficiently utilize the vacant positions in a CPI-SCMA codeword.

In HCPI-SCMA, unlike CPI-SCMA, the total available positions are divided into $R$ parts, where $R$ is the hybrid order (HO). Therefore, 
\begin{equation}
m = \sum_{r=1}^{R} m_1^{(r)} + m_2^{(r)}.
\end{equation}
The number of bits that are used to convey data in CPI-SCMA can directly extend to HCPI-SCMA, i.e.,
\begin{equation}
m_2^{(r)} = \sum_{r=1}^{R} t^{(r)}\log_2 C. \label{HCPI_m2}
\end{equation}
It can be observed from \eqref{HCPI_m2} that $t = \sum_{r=1}^{R} t^{(r)}$. As for $m_1^{(r)}$ in HCPI-SCMA, it should be modified as:
\begin{equation}
m_1^{(r)} = \sum_{r=1}^{R} \left\lfloor \log_2\left(C(n^{(r)}-\sum_{\gamma=0}^{r-1}n^{\gamma},t^{(r)})\right)\right \rfloor.
\end{equation}
Note that each HO corresponds to a different codebook $\mathcal{S}_r$. Therefore, the $r$th order transmitted vector for user $j$ is:
\begin{equation}
\boldsymbol X_j^{(r)} =  \{\boldsymbol{x}_{j,c_1}^{(r)}, \boldsymbol{x}_{j,c_2}^{(r)}, \cdots \boldsymbol{x}_{j,c_{t^{(r)}}}^{(r)}\}. \label{signal_space}
\end{equation}
Furthermore, the signal space of all users is:
\begin{equation}
\mathbb{X} = \bigcup_{j=1}^J\bigcup_{r=1}^R \boldsymbol X_j^{(r)}.
\end{equation}
Moreover, the index space of each HCPI-SCMA codeword can be represented as:
\begin{equation}
\mathcal{I} = \bigcup_{r=1}^R \{i_{1}, i_{2}, \cdots, i_{t^{(r)}}\}, \label{index_space}
\end{equation}
where $i_{\beta} \in [1, \cdots, n]$, $\beta \in [1, \cdots, t^{(r)}]$, and $\boldsymbol{x}_{j,c_{\beta}}^{(r)} \in \mathcal{S}_r$. It should be noted that the index space are both unitary for each user, which indicates that each user shares the same pattern to construct a HCPI-SCMA codeword. After constructing the codewords, they are sent to Rayleigh fading channel, and superimposed at the receiver. Therefore, the received signals of HCPI-SCMA is:
\begin{equation}
\boldsymbol y = \sum_{j=1}^J diag(\boldsymbol h_j)\boldsymbol c_j + \boldsymbol z, \label{superpose_sig}
\end{equation}
where $diag(\boldsymbol h_j)$ is the equalized channel matrix, $\boldsymbol h_j = [h_1^j, \cdots, h_s^j, \cdots, h_S^j]$. $\boldsymbol z$ is the noise vector that follows complex Gaussian distribution $\mathcal{CN}(0,\sigma^2\boldsymbol{I})$; $\boldsymbol{I}$ is an identity vector and $\boldsymbol y = [ y_1, \cdots,y_{s} ,\cdots, y_{S}]$.

\subsection{HCPI-SCMA codeword construction}
To construct a HCPI-SCMA codeword $\boldsymbol{c}_j$, $R$ order index modulation should be operated. As for the first order index modulation, the process is the same as CPI-SCMA, i.e., the activated codeword positions are decided according to the BTI mapper as shown in Table \ref{LUT}. Note that the mappings between bits and indices in Table \ref{LUT} is only one example, which means that different BTI mapper can be employed. 
\begin{table}[h]\scriptsize
	\renewcommand{\arraystretch}{1.3}
	\caption{A BTI mapper example for $n = 4, t = 2$}
	\label{LUT}
	\centering
	\begin{tabular}{|p{0.8cm}||p{0.8cm}||p{2.0cm}|}
		\hline
		bits &  indices & data block  \\\hline
		
		$[0\quad0]$ & $\{1 \quad3\}$ & [$\boldsymbol{x}_{j,c_1}$ $\boldsymbol{0}$ $\boldsymbol{x}_{j,c_2}$ $\boldsymbol{0}$] \\\hline
		
		$[0\quad 1]$ & $\{2 \quad4\}$ & [$\boldsymbol{0}$ $\boldsymbol{x}_{j,c_1}$  $\boldsymbol{0}$ $\boldsymbol{x}_{j,c_2}$]\\\hline
		
		$[1 \quad0]$ & $\{2\quad 3\}$ & [$\boldsymbol{0}$ $\boldsymbol{x}_{j,c_1}$ $\boldsymbol{x}_{j,c_2}$ $\boldsymbol{0}$] \\\hline
		
		$[1 \quad1]$ & $\{1\quad 4\}$ & [$\boldsymbol{x}_{j,c_1}$ $\boldsymbol{0}$  $\boldsymbol{0}$ $\boldsymbol{x}_{j,c_2}$] \\\hline
	\end{tabular}
\end{table}

For the transmitted of HCPI-SCMA, firstly, all the positions are used to process bits to indices mapping shown in Table \ref{LUT}, and the $t$ active positions are occupied with codewords in the first codebook. After this point, the remainder of available positions are occupied with SCMA codewords selected from another codebooks, and their corresponding indices are modulated by a new BTI mapper. Such operations continue until reaching the maximum HO.

For simplicity, a HCPI-SCMA codeword construction with $R = 2, n = n^{(1)} = 4, t^{(1)}=2, t^{(2)}=1$ is taken as an example in this paper.
\begin{table}[t!]\scriptsize
	\renewcommand{\arraystretch}{1.3}
	\caption{A BTI mapper example for HCPI-SCMA with $R = 2, n = n^{(1)} = 4, t^{(1)}=2, t^{(2)}=1$}
	\label{HCPI_LUT}
	\centering
	\begin{tabular}{|p{1.1cm}||p{1.4cm}||p{2.5cm}|}
		\hline
		bits &  indices & data block  \\\hline
		
		$[0\quad0], [0]$ & $\{1 \quad3\}, \{2\}$ & [$\boldsymbol{x}_{j,c_1}^{(1)}$ $\boldsymbol{x}_{j,c_1}^{(2)}$ $\boldsymbol{x}_{j,c_2}^{(1)}$ $\boldsymbol{0}$] \\\hline
		
		$[0\quad0], [1]$ & $\{1 \quad3\}, \{4\}$ & [$\boldsymbol{x}_{j,c_1}^{(1)}$ $\boldsymbol{0}$ $\boldsymbol{x}_{j,c_2}^{(1)}$ $\boldsymbol{x}_{j,c_1}^{(2)} $] \\\hline
		
		$[0\quad 1], [0]$ & $\{2 \quad4\}, \{1\}$ & [$\boldsymbol{x}_{j,c_1}^{(2)}$ $\boldsymbol{x}_{j,c_1}^{(1)}$  $\boldsymbol{0}$ $\boldsymbol{x}_{j,c_2}^{(1)}$]\\\hline
		
		$[0\quad 1], [1]$ & $\{2 \quad4\}, \{3\}$ & [$\boldsymbol{0}$ $\boldsymbol{x}_{j,c_1}^{(1)}$  $\boldsymbol{x}_{j,c_1}^{(2)}$ $\boldsymbol{x}_{j,c_2}^{(1)}$]\\\hline
		
		$[1 \quad0], [0]$ & $\{2\quad 3\}, \{1\}$ & [$\boldsymbol{x}_{j,c_1}^{(2)}$ $\boldsymbol{x}_{j,c_1}^{(1)}$ $\boldsymbol{x}_{j,c_2}^{(1)}$ $\boldsymbol{0}$] \\\hline
		
		$[1 \quad0], [1]$ & $\{2\quad 3\}, \{4\}$ & [$\boldsymbol{0}$ $\boldsymbol{x}_{j,c_1}^{(1)}$ $\boldsymbol{x}_{j,c_2}^{(1)}$ $\boldsymbol{x}_{j,c_1}^{(2)}$] \\\hline
		
		$[1 \quad1], [0]$ & $\{1\quad 4\}, \{2\}$ & [$\boldsymbol{x}_{j,c_1}^{(1)}$ $\boldsymbol{x}_{j,c_1}^{(2)}$  $\boldsymbol{0}$ $\boldsymbol{x}_{j,c_2}^{(1)}$] \\\hline
		
		$[1 \quad1], [1]$ & $\{1\quad 4\}, \{3\}$ & [$\boldsymbol{x}_{j,c_1}^{(1)}$ $\boldsymbol{0}$  $\boldsymbol{x}_{j,c_1}^{(2)}$ $\boldsymbol{x}_{j,c_2}^{(1)}$] \\\hline
	\end{tabular}
\end{table}
As shown in Table \ref{HCPI_LUT}, the first $m_1^{(1)} = 2$ bits (the left hand side in the first column) select $t^{(1)}=2$ active positions to place SCMA codewords in $\mathcal{S}_1$. As $n-t^{(1)}$ positions are still accessible, an extra bit (the right hand side in the first column) is utilized to select which position can be occupied with an SCMA codeword chosen from $\mathcal{S}_2$. By employing such a hybrid index modulation in CPI-SCMA, an HCPI-SCMA codeword $\boldsymbol{c_j}$ consists of $t^{(1)} + t^{(2)} = 3$ $K$-dimensional signal vectors and one all-zero vector $\boldsymbol{0}$ is constructed. As can also be observed from the table, in the second order index modulation, 0 maps to the first vacant position while 1 maps to the last one after the first order index modulated codewords are generated.

Similarly, the HCPI-SCMA codewords are transmitted over the channel and then superimposed at the receiver, and the received signals are formed as shown in \eqref{superpose_sig}. In contrast to CPI-SCMA, multiple BTI mappers and SCMA codebooks are adopted to construct an HCPI-SCMA codeword, therefore, a new detection algorithm of HCPI-SCMA should be provided.
\section{Detection Algorithm}
In this section, a novel algorithm that is used to detect the conveyed information of HCPI-SCMA is introduced.
\subsection{Merging codebook based MPA }
In HCPI-SCMA, the transmitted signals are encoded by $R$ different $C$-point SCMA codebooks, which indicates that codewords selected from different codebooks may be superimposed in one detection unit \footnote{In both C-SCMA and CPI-SCMA, one detection unit equals to $K$ received chips.}. To detect the conveyed information, a merging codebook based MPA (MC-MPA) is proposed in this paper.

As discussed above, there are $S$ complex symbols at the receiver, note that each chip is a superimposed signal consists of symbols selected from diverse $\mathcal{S}_r$ and zero. Consequently, different codebooks can be merged into one codebook, which indicates that the combination set in the conventional SCMA should be modified as:
\begin{equation}
	\chi_{hcpi} = \left[\mathcal{X}_{i_1}^1 \cdots \mathcal{X}_{i_1}^R  \quad \boldsymbol{0}\right] \times \cdots \times \left[\mathcal{X}_{i_{d_f}}^1 \cdots \mathcal{X}_{i_{d_f}}^R \quad \boldsymbol{0}\right],
\end{equation}
where $\times$ denotes the Cartesian product. Then the MC-MPA uses the new combination set to update function node (FN) in a detection unit can be rewritten as: 
\begin{equation}
I_{f_k\to v_j}(\boldsymbol{x}_{j,c}) = \sum\limits_{\Delta\in \chi_{hcpi}}\dfrac{1}{\pi N_0}\exp(d_{k,j})
\prod\limits_{i\in \xi_k/j}I_{v_i\to f_k}\left(L_i\right),\label{FN_update}
\end{equation}
where $\xi_k$ is a node set consists of all user nodes (UNs) that connect to $k$th FN,  $L_{i_1}=\{l_k^i\}_{k=1}^K$, and 
$d_{k,j}$ can be further written as:
\begin{equation}
	d_{k,j} = -\dfrac{1}{N_0}\left\Vert y_k-h_k^j x_{c,k}^j-\sum\limits_{i\in \xi_k/j}h_k^il_k^i\right\Vert ^2. \label{Euc}
\end{equation}

The following steps in MC-MPA are similar to the original MPA in SCMA. However, $(RC+1)$ dimensional messages are exchanged between FN and UN in the Tanner graph rather than $C$ dimension in HCPI-SCMA. It should be noted that the reliabilities are defined in the probability domain in this paper, and the obtained  messages of each user are normalized to unity.
Furthermore, since different codebooks are utilized in HCPI-SCMA, the design of used codebooks are of significant importance. This follows from the fact that once the Euclidean distances among these codebooks become close, they will be difficult to be distinguished from each other, leading to decision errors. Therefore, the design of adopted codebooks in this scheme has significant impact on the error rate performance of HCPI-SCMA.

At the HCPI-SCMA receiver, each detection unit processes MC-MPA in sequence, note that the component of a detection unit is merely a part of the whole HCPI-SCMA codeword. Consequently, MC-MPA cannot obtain the real soft messages of transmitted bits in a HCPI-SCMA codeword. To obtain the real soft information, a joint algorithm capable of detecting $S$ detection units at once should be designed. As such, the HCPI-SCMA can be combined with error control codes in a straightforward manner.

\subsection{MC-MPA aided detection algorithm}
After processing MC-MPA in HCPI-SCMA, the obtained messages are still insufficient to detect the indices and data; hence, the message vectors that contain the reliabilities of each symbol in a HCPI-SCMA codeword can be acquired. To recover the indices and their corresponding data, further operations are needed to handle those messages.

For simplicity, we define a $J\dots (RC + 1)$ detected message matrix for detection unit $\kappa$ as:
\begin{equation}
\begin{array}{l}
\boldsymbol{\phi}_{\kappa}=
\left[
\begin{matrix}
p_{1,1}^{\kappa}&p_{1,2}^{\kappa}&\cdots&\cdots&p_{1,RC}^{\kappa}&p_{1,0}^{\kappa}\\
p_{2,1}^{\kappa}&p_{2,2}^{\kappa}&\cdots&\cdots&p_{2,RC}^{\kappa}&p_{2,0}^{\kappa}\\
\vdots&\vdots&\vdots&\vdots&\vdots&\vdots\\
p_{j,1}^{\kappa}&p_{j,2}^{\kappa}&\cdots&\cdots&p_{j,RC}^{\kappa}&p_{j,0}^{\kappa}\\
\vdots&\vdots&\vdots&\vdots&\vdots&\vdots\\
p_{J,1}^{\kappa}&p_{J,2}^{\kappa}&\cdots&\cdots&p_{J,RC}^{\kappa}&p_{J,0}^{\kappa}
\label{soft_information_matrix}
\end{matrix}
\right],
\end{array}
\end{equation}
where $\kappa \in \{1, \cdots, n\}$, and $p_{j,q}^{\kappa}$ is the probability domain message that represents the reliability of the $q$th ($q \in \{0,1,\cdots, RC\}$) symbol in the $\kappa$th codeword position of user $j$.

Taking the HCPI-SCMA system in Table \ref{HCPI_LUT} as an example, the MC-MPA aided detection of HCPI-SCMA can be divided into following three steps:

\subsubsection{Detection of unoccupied positions}

Since the inactive codeword positions are padded with zeros, and the Euclidean distance between $\boldsymbol{0}$
and an arbitrary SCMA codeword equals to 1 as long as the power of C-SCMA codebook is normalized to unity. Moreover, in HCPI-SCMA, unoccupied positions are much less than occupied positions. It is thus more likely to make correct decisions on the vacant positions. Consequently, they are detected at first, which can be represented as:
\begin{equation}
	\mathbf{I}_0^j = \arg \max_{\kappa = 1,\cdots, n} \left[(p_{j,0}^{\kappa}), (n-\sum_{r=1}^{R}t^{(r)})\right], \label{index_0}
\end{equation}
where $\max [(i),(j)]$ denotes selecting $j$ maximum elements from set $i$， and the set $\mathbf{I}_0^j$ contains $j$th user's indices of unoccupied positions. It should be noted that the erroneous detection of vacant positions may lead to catastrophic results since the errors will propagate to the following detection. From this perspective, this step decide the performance of HCPI-SCMA to a large extent.

\subsubsection{First order index modulation detection}
After the detection of vacant codeword indices, the number of possible active positions in the first order index modulation  decreases. in Table \ref{LUT}, if the third position is detected as unoccupied for $j$th user, then only the second and forth mappings in the table should be considered. To judge which case is more likely to happen, we consider the following probability:
\begin{equation}
	P_{\kappa}^j = \sum_{q=1}^{M} p_{j,q}^{\kappa}, \label{pro_codenook}
\end{equation}
where $P_{\kappa}^j$ is the probability that $j$th user adopts the codebook $\mathcal{S}_1$ in the $\kappa$th position. To make a joint decision, we consider the joint probability of all candidate indices, i.e.,
\begin{equation}
    \mathbf{I}_1^j = \arg \max_{\kappa \in \mathbf{I}_c^{(1)}  } \prod_{\kappa} P_{\kappa}^j, \label{index_1}
\end{equation}
where $\mathbf{I}_c^{(1)}$ consists of all possible index combinations when the unoccupied positions are known. After this point, the corresponding data at the positions include in $\mathbf{I}_1^{j}$ is detected according to maximum likelihood rule, i.e.,
\begin{equation}
\mathbf{D}_1^j = \arg \max_{q = 1,\cdots, C} \phi_{\kappa}(j,q) \quad \kappa \in \mathbf{I}_1^j. \label{data_1}
\end{equation}
According to \eqref{index_0}, \eqref{index_1} and \eqref{data_1}, the transmitted bits that process first order index modulation can be recovered.
\subsubsection{High order index modulation detection}
 As for the second order or higher order index modulation, the detection process is similar to the first oder detection. However, the cardinality of candidate set $\mathbf{I}_c^{(r)}$ that contains all possible index combinations in the $r$th order detection decreases, which indicates that less operations are necessary for recovering the data with high order index modulation. By combining the index set $\mathbf{I}_r^j$ and data set $\mathbf{D}_r^j$ ($r = 1, \cdots R$) of each order, the information bits that construct a HCPI-SCMA codeword can be obtained. It should be noted that once the first oder information cannot be de-modulated without errors, they will certainly propagate to the following detection, leading to catastrophic results.
For simplicity, MPAD can be summarized as Algorithm 1:

\begin{algorithm}[h]	
	\caption{Detection algorithm for HCPI-SCMA}	
	\begin{algorithmic}[1]  
		\STATE  \textbf{MC-MPA:}\\
		\FOR {$\text{iter}=1,\cdots,\text{iter\_num}$}	
		\STATE \textbf{Initialization:} \\
		\quad\quad\quad\quad\quad\quad\quad$I_{v_j\to f_k}(\boldsymbol{x}_{j,c}) = 1/C$

		\STATE \textbf{Function Node (FN) update:} \\	
		Using $\eqref{FN_update}$ and $\eqref{Euc}$ to update FN
		\\

		\STATE \textbf{User Node (UN) update:} \\	
		\quad\quad\quad\quad$I_{v_j \to f_k}(\boldsymbol{x}_{j,c}) = \prod\limits_{l \in \zeta_j/k} I_{f_l \to v_j}(\boldsymbol{x}_{j,c})$,\\
		where $\zeta_j$ is a node set contains FNs that connect to $j$th UN.
		
		\STATE  \textbf{Decision:}\\
		\quad\quad\quad\quad$I_{v_j \to f_k}(\boldsymbol{x}_{j,c}) = \prod\limits_{k\in \zeta_j}I_{f_k \to v_j}(\boldsymbol{x}_{j,c})$
		
		\ENDFOR
		
		\STATE  \textbf{Detection of unoccupied positions:}\\
		1. Constructing the matrix in $\eqref{soft_information_matrix}$ for each detection unit.\\
		2. Using \eqref{index_0} to detect the vacant codeword position indices, and obtain $\mathbf{I}_0^j$.
		
		\STATE  \textbf{First order index modulation detection:}\\
		1. Calculating the probabilities of the using codebook on the activated positions by processing $\eqref{pro_codenook}$ and $\eqref{index_1}$.\\
		2. Making decision on the symbol of each active codewprd positions by using $\eqref{data_1}$.
		\STATE  \textbf{High order index modulation detection:}\\
		1. Calculating the $r$th order candidate indices set $\mathbf{I}_c^{(r)}$,  where $r = 1, \cdots R$.\\
		2. Repeating step 1 and 2 in the first order index modulation detection until reaching the maximum HO, and obtain the index set $\mathbf{I}_r^j$ and data set $\mathbf{D}_r^j$. 
		\STATE  \textbf{Final decision:}\\
		Combining the results of $\mathbf{I}_r^j$ and $\mathbf{D}_r^j$, where $r = 0, \cdots R$.
		
	\end{algorithmic}
\end{algorithm}

It can be inferred that the computational complexity of MC-MPA reaches to $O((RC+1)^{d_f})$. From above discussions, many extra operations need to be carried out upon the completion of MC-MPA to recover the transmitted bits; hence, the complexity of HCPI-SCMA is higher than C-SCMA with complexity order $O(C^{d_f})$ when the HO is high. From this perspective, there exists a trade-off between the improvement of TE and computational complexity, which indicates that a higher TE is achieved by addind complexity in HCPI-SCMA system.

\section{Numerical results and analysis}	
In this section, the analysis and simulation results are presented to demonstrate the effectiveness of our proposed scheme.

\subsection{Error rate performance analysis}
To analyze the error rate performance of HCPI-SCMA, the average block error probability (ABLEP) is estimated. Considering a HCPI-SCMA system with HO equals to $R$, we first define the possible transmitted signals of all users as
$\mathbf{C} = \left[\boldsymbol{c}_1,\cdots,\boldsymbol{c}_j,\cdots,\boldsymbol{c}_J \right]$.
As shown in \cite{Tse2009Fundamentals}, the conditional pairwise error probability of such model can be written as:
\begin{equation}
P(\mathbf{C} \to \mathbf{\hat{C}}|\mathbf{h}) = Q \left(\sqrt{\dfrac{\mathbf{h}^{\dagger}\Delta\Delta^H\mathbf{h}}{2N_0}}\right),\label{CPEP}
\end{equation}
where $Q$ is the Q-function, 
$\mathbf{\hat{C}}$ is the erroneously detected candidate codeword set, and $\Delta=\Vert (\mathbf{C}-\mathbf{\hat{C}})\Vert^{2} = (\mathbf{C}-\mathbf{\hat{C}})^{H}(\mathbf{C}-\mathbf{\hat{C}})$ \footnote{$[\cdot]^H$ and $[\cdot]^{\dagger}$ denotes the Hermitian operator and conjugate transpose, respectively.}. Furthermore, the Q-function can be approximated by
$Q(x) \approx 1/12e^{-x^2/2}+1/4e^{-2x^2/3}$.
Therefore, the unconditional pairwise error probability (UPEP) can be calculated by operating mathematical expectation on \eqref{CPEP}.
Since $\Delta\Delta^H$ is hermitian, which indicates that it can be diagonalized as $\Delta \Delta^{H} = U diag\{\lambda_1^2, \lambda_2^2, \cdots, \lambda_{S}^2, 0, \cdots, 0\}U^{H}$, where $U$ is a unitary matrix, and $\lambda_{s}^2$ are the non-zero eigenvalue of $\Delta\Delta^H$, which can be further written as:
\begin{equation}
\lambda_{s}^2 = \sum_{j=1}^J\vert c_j[s] -  \hat{c}_j[s]\vert^2.
\end{equation}
By using the approximation of Q-function
Note that $c_j[s] \in \bigcup_{r=1}^R\mathcal{S}_r$, which indicates that $K^{r+1}$ different symbols can be selected for each user. The unconditional pairwise error probability can be represented as:
\begin{equation}
	P(\mathbf{C} \to \mathbf{\hat{C}}) \approx \prod_{s=1}^{S} \left(\dfrac{1}{12}\dfrac{2N_0}{2N_0-1/2\lambda_s^2} + \dfrac{1}{4}\dfrac{2N_0}{2N_0-2/3\lambda_s^2}\right). \label{UPEP}
\end{equation}

Following the approach in \cite{Zhu2002Performance}, the ABLEP for the $j$th user by using MLD is upper bounded by:
\begin{equation}
P_j(e) \leq \dfrac{1}{(\prod_{r=1}^R2^{m_1^{(r)}}C^{t^{(r)}})^J} \sum_{\mathbf{C}} \left( \sum_{\mathbf{\hat{C}},\mathbf{c^j \neq \hat{c}^j}} P(\mathbf{C} \to \mathbf{\hat{C}})\right). \label{ABLER}
\end{equation}
By combining \eqref{UPEP} and \eqref{ABLER}, ABLEP of HCPI-SCMA can be calculated.
\subsection{BER and TE comparison}
The simulation parameters are set to  $J = 6;K = 4;C = 4$, the BTI mapper for CPI-SCMA and HCPI-SCMA are shown in Table \ref{LUT} and Table \ref{HCPI_LUT}, respectively. The codebook used for C-SCMA and CPI-SCMA are designed according to \cite{Taherzadeh2014SCMA,Dai2017Partially,Cai2016Multi}, which has proved to achieve satisfactory performance in the literature. As for HCPI-SCMA, the adopted codebooks are selected from these three kinds of codebook.

To make fair comparison, we first define TE. In general, TE indicates the ability to transmit bits per physical resource (chip), and thus the TE of CPI-SCMA is represented as:
\begin{equation}
r_t^{cpi}= \dfrac{J(m_1 + m_2)}{nK}. \quad \text{(bits/chip)}
\end{equation}
Moreover, TE of SCMA $r_t^{c}$ equals to  $(J\log_2C)/K$.
It can be observed that once the indicator matrix and size of codebook is determined, $r_t^{c}$ is a fixed value. For instance, as for a C-SCMA system with $J=6, K=4, C=4$, the TE of it is 3 bits/chip.

For HCPI-SCMA, we first define $n^0 = 0, n^1 = n$. It should be noted that the performance comparison is based on such fact that CPI-SCMA and HCPI-SCMA both use the same number of physical resources. Therefore, TE in HCPI-SCMA can be represented as:
\begin{equation}
\begin{aligned}
r_t^{hcpi} &= \dfrac{J(m_1 + m_2)}{nK}\\
&=\dfrac{\sum\limits_{r=1}^{R} \left[\lfloor \log_2(C(n^{(r)}-\sum\limits_{\gamma=0}^{r-1}n^{(\gamma)},t^{(r)})) \rfloor + t^{r}\log_2 C \right]}{nK} \\
C \label{TE_HCPI}
\end{aligned}
\end{equation}
Note that \eqref{TE_HCPI} is a union formation of TE for both CPI-SCMA and HCPI-SCMA. When $R=1$, TE of HCPI-SCMA degenerates to CPI-SCMA, and thus the $r_t^{hcpi} = r_t^{cpi}$. Once $R \geq 1$, the inequality in \eqref{TE_HCPI} is satisfied. Therefore, it can be inferred that HCPI-SCMA can always achieve higher TE than CPI-SCMA under the condition that they both have the same number of resource blocks.

\begin{figure}[h]
	\centering
	
	\renewcommand{\captionfont}{\small } \renewcommand{\captionlabelfont}{\small} \centering \includegraphics[height=3in, width=3.5in]{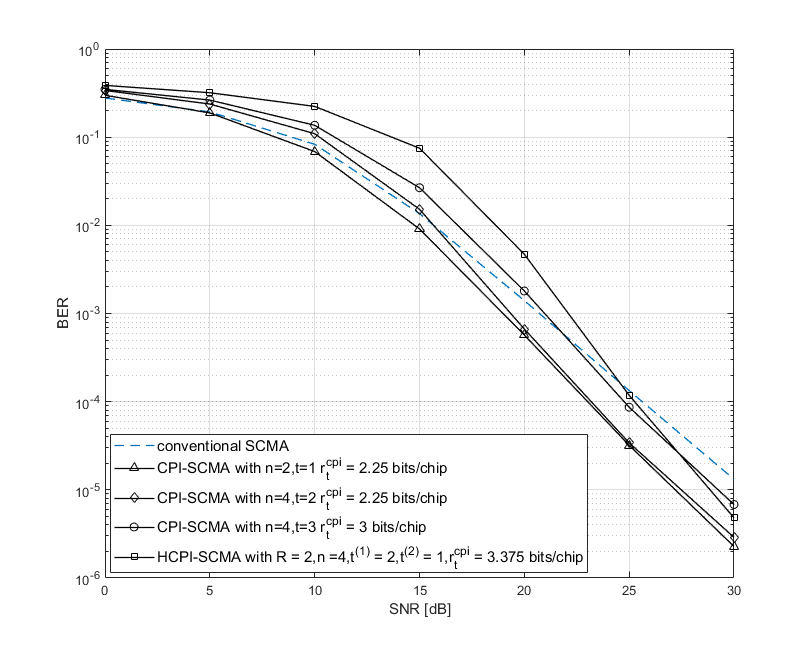}\caption{Performance Comparison of C-SCMA, CPI-SCMA and HCPI-SCMA with $n=4$.}
	
	\label{fig_HCPI_CPI_comp_n_4}
	
\end{figure}
In Fig. \ref{fig_HCPI_CPI_comp_n_4}, the error rate performance of HCPI-SCMA system with $R = 2, n = n^{(1)} = 4, t^{(1)}=2, t^{(2)}=1$ is compared with C-SCMA and different CPI-SCMA systems. As can be observed from the figure, the performances of CPI-SCMA, which is equivalent to HCPI-SCMA with single index modulation, i.e., CPI-SCMA with $n=4, t=2$ and $n=2, t=1$ in the figure, are better than HCPI-SCMA at all SNRs, especially in the region of low SNR. CPI-SCMA with $n=4, t=2$ and $n=2, t=1$ can both achieve 3 dB gain at BER = $10^{-3}$. However, their TE  far lower than that of HCPI-SCMA. As for CPI-SCMA with $n=4, t=3$, which has the same number of unoccupied positions as HCPI-SCMA simulated in the figure. Furthermore, HCPI-SCMA suffers from at most 4 dB performance loss in the low SNR region, which can be explained by the error propagation phenomenon in HCPI-SCMA, i.e., once the indices of first order modulation are detected with error, the erroneous results can affect the following detection. Nonetheless, the curve of HCPI-SCMA descends dramatically when SNR is greater than 15 dB, which indicates that the slope of it is even steeper than CPI-SCMA with $n=2, t=1$ and $n=4, t=2$. Hence, HCPI-SCMA outperforms CPI-SCMA with $n=4, t=3$ at high SNRs. Moreover, both HCPI-SCMA and CPI-SCMA in the figure perform better than C-SCMA in the region of high SNR.

\begin{figure}[h]
	\centering
	
	\renewcommand{\captionfont}{\small } \renewcommand{\captionlabelfont}{\small} \centering \includegraphics[height=3in, width=3.5in]{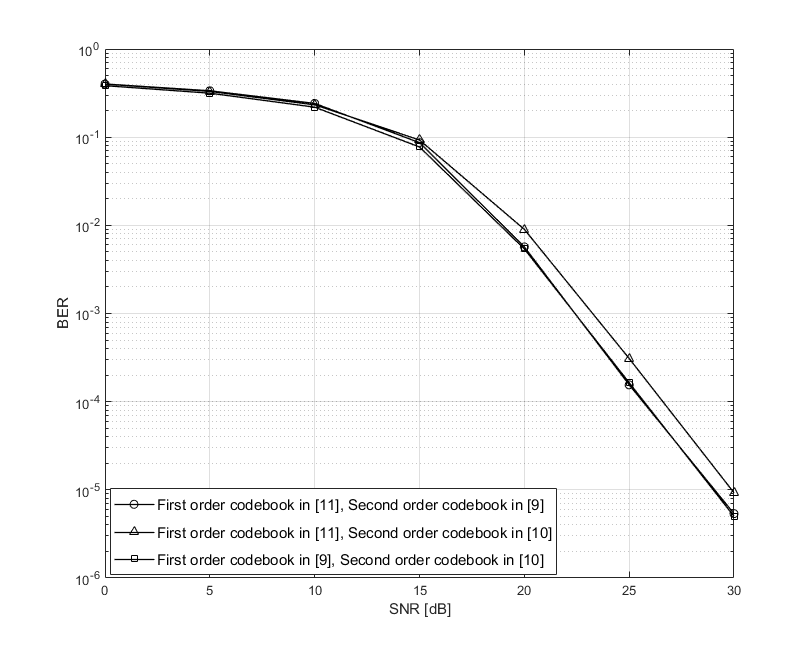}\caption{Performance Comparison of different codebook combinations.}
	
	\label{fig_HCPI_CPI_diff_cb}
	
\end{figure}

To show the effects of codebook design in HCPI-SCMA, the error rate performances of HCPI-SCMA with different codebook combinations are presented in Fig. \ref{fig_HCPI_CPI_diff_cb}. Note that the HCPI-SCMA system under investigation is the same as in Fig. \ref{fig_HCPI_CPI_comp_n_4}. As can be seen from the figure, the codebook selection can affect the error rate performance, especially when SNR is greater than15 dB. This is due to the fact that a merging codebook is utilized in the MPA, which indicates that SCMA codewords in different codebooks are involved in the MC-MPA. Since the messages obtained from MC-MPA will be further used to detect the information carried by indices and their corresponding data, all the used codebooks can  affect the error rate performance. Therefore, the codebook design in the HCPI-SCMA is of significant importance.

\begin{figure}[h]
	\centering
	
	\renewcommand{\captionfont}{\small } \renewcommand{\captionlabelfont}{\small} \centering \includegraphics[height=3in, width=3.5in]{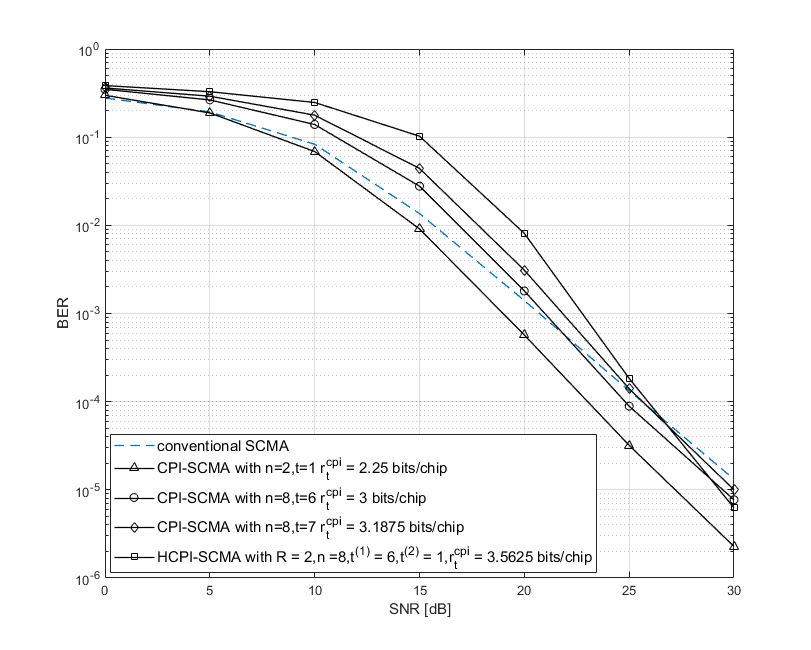}\caption{Performance Comparison of C-SCMA, CPI-SCMA and HCPI-SCMA with $n=8$.}
	
	\label{fig_HCPI_CPI_comp_n_8}
	
\end{figure}

The HCPI-SCMA with $R =2, n=n^{(1)}=8, t^{(1)}=6, t^{(2)}=1$ is also simulated, and performance shown in Fig. 3 where from Fig. \ref{fig_HCPI_CPI_comp_n_8}, the error rate performance of HCPI-SCMA with more available codeword positions is presented. In general, the simulation results are similar to the HCPI-SCMA with $R =2, n=n^{(1)}=4, t^{(1)}=2, t^{(2)}=1$. It is shown in the figure that the performance of HCPI-SCMA also deteriorates significantly when SNR is less than 15 dB, however, it can converge to the performance of CPI-SCMA with $n=8, t=6$ as SNR increasing. The proposed HCPI-SCMA provides a new method to design SCMA system with high flexibility, which is widely applicable to different scenarios. Furthermore, in practical systems, the multiple access system can be  designed to be TE adaptable by using CPI-SCMA and HCPI-SCMA.

\section{Conclusion}	
In this paper, we proposed a novel HCPI-SCMA system employing hybrid index modulation in a codeword to obtain higher TE compared with CPI-SCMA. Since hybrid BTI mappers and codebooks are utilized in the HCPI-SCMA system, a novel detection algorithm utilizing the merging codebook was proposed to detect the activated indices which adopt different BTI mappers and their corresponding data.

\ifCLASSOPTIONcaptionsoff
\newpage
\fi
\bibliographystyle{IEEEtran}
\bibliography{Hybrid_CPI_SCMA_ICC}

\begin{thebibliography}{10}
\providecommand{\url}[1]{#1}
\csname url@samestyle\endcsname
\providecommand{\newblock}{\relax}
\providecommand{\bibinfo}[2]{#2}
\providecommand{\BIBentrySTDinterwordspacing}{\spaceskip=0pt\relax}
\providecommand{\BIBentryALTinterwordstretchfactor}{4}
\providecommand{\BIBentryALTinterwordspacing}{\spaceskip=\fontdimen2\font plus
\BIBentryALTinterwordstretchfactor\fontdimen3\font minus
  \fontdimen4\font\relax}
\providecommand{\BIBforeignlanguage}[2]{{%
\expandafter\ifx\csname l@#1\endcsname\relax
\typeout{** WARNING: IEEEtran.bst: No hyphenation pattern has been}%
\typeout{** loaded for the language `#1'. Using the pattern for}%
\typeout{** the default language instead.}%
\else
\language=\csname l@#1\endcsname
\fi
#2}}
\providecommand{\BIBdecl}{\relax}
\BIBdecl

\bibitem{Andrews2014What}
J.~G. Andrews, S.~Buzzi, C.~Wan, and S.~V. Hanly, ``What will 5{G} be?''
  \emph{IEEE Journal on Selected Areas in Communications}, vol.~32, no.~6, pp.
  1065--1082, Jun. 2014.

\bibitem{Hosein2013SCMA}
H.~Nikopour and H.~Baligh, ``Sparse code multiple access,'' in \emph{Proc. IEEE
  24th Int. Symp. Pers. Indoor Mobile Radio Commun. (PIMRC)}, London, U.K.,
  Oct. 2013, pp. 332--336.

\bibitem{Mesleh2008SM}
R.~Y. Mesleh, H.~Haas, S.~Sinanovic, W.~A. Chang, and S.~Yun, ``Spatial
  modulation,'' \emph{IEEE Transactions on Vehicular Technology}, vol.~57,
  no.~4, pp. 2228--2241, Jul. 2014.

\bibitem{Yang2014Design}
P.~Yang, M.~D. Renzo, Y.~Xiao, and S.~Li, ``Design guidelines for spatial
  modulation,'' \emph{IEEE Communications Surveys and Tutorials}, vol.~17,
  no.~1, pp. 6--26, Furst quarter 2015.

\bibitem{Ba2013Orthogonal}
E.~Başar, Ümit Aygölü, E.~Panayırcı, and H.~V. Poor, ``Orthogonal
  frequency division multiplexing with index modulation,'' \emph{IEEE
  Transactions on Signal Processing}, vol.~61, no.~22, pp. 5536--5549, Nov.
  2013.

\bibitem{Dang2018adaptive}
S.~Dang, J.~P. Coon, and G.~Chen, ``Adaptive {O}{F}{D}{M} with index modulation
  for two-hop relay-assisted networks,'' \emph{IEEE Transactions on Wireless
  Communications}, vol.~17, no.~3, pp. 1923--1936, Mar 2018.

\bibitem{LaiCPI}
K.~Lai, L.~Wen, L.~Jing, G.~Chen, P.~Xiao, and M.~Amine, ``Codeword position
  index based sparse code multiple access system,'' \emph{arXiv preprint
  arXiv:1811.03777}, 2018.

\bibitem{Liu2017Spatial}
Y.~Liu, L.~L. Yang, and L.~Hanzo, ``Spatial modulation aided sparse
  code-division multiple access,'' \emph{IEEE Transactions on Wireless
  Communications}, vol.~17, no.~3, pp. 1474--1487, Mar. 2018.

\bibitem{Zhu2017NOMA}
X.~Zhu, Z.~Wang, and J.~Cao, ``{N}{O}{M}{A}-based spatial modulation,''
  \emph{IEEE Access}, vol.~5, no.~99, pp. 3790--3800, Apr 2017.

\bibitem{Tse2009Fundamentals}
D.~Tse and P.~Viswanath, \emph{Fundamentals of wireless communication}.\hskip
  1em plus 0.5em minus 0.4em\relax Cambridge, U.K, Cambridge Univ. Press, 2005.

\bibitem{Zhu2002Performance}
X.~Zhu and D.~Murch, R, ``Performance analysis of maximum likelihood detection
  in a {M}{I}{M}{O} antenna system,'' \emph{IEEE Transactions on
  Communications}, vol.~50, no.~2, pp. 187--191, Feb. 2002.

\bibitem{Taherzadeh2014SCMA}
M.~Taherzadeh, H.~Nikopour, A.~Bayesteh, and H.~Baligh, ``{S}{C}{M}{A} codebook
  design,'' in \emph{Proc. IEEE Veh. Technol. Conf. (VTC-Fall)}, Vancouver,
  Canada., Sep. 2014, pp. 1--5.

\bibitem{Dai2017Partially}
J.~Dai, G.~Chen, K.~Niu, and J.~Lin, ``Partially active message passing
  receiver for {M}{I}{M}{O}-{S}{C}{M}{A} systems,'' \emph{IEEE Wireless
  Communications Letters}, vol.~7, no.~2, pp. 222--225, Apr. 2018.

\bibitem{Cai2016Multi}
D.~Cai, P.~Fan, X.~Lei, Y.~Liu, and D.~Chen, ``Multi-dimensional {S}{C}{M}{A}
  codebook design based on constellation rotation and interleaving,'' in
  \emph{Proc. IEEE Veh. Technol. Conf. (VTC-Fall)}, Nanjing, China, May. 2016,
  pp. 1--5.

\end{thebibliography}

\end{document}